\documentclass[a4paper,11pt]{article}

\usepackage{jheppub} 

\usepackage[T1]{fontenc} 

\usepackage{caption} 
\usepackage{slashed} 
\usepackage{verbatim} 
\usepackage{bbold}
\usepackage{braket}
\usepackage[compat=1.1.0]{tikz-feynman}
\usepackage{ytableau}

\def\simlt{\mathrel{\lower2.5pt\vbox{\lineskip=0pt\baselineskip=0pt
           \hbox{$<$}\hbox{$\sim$}}}}
\def\simgt{\mathrel{\lower2.5pt\vbox{\lineskip=0pt\baselineskip=0pt
           \hbox{$>$}\hbox{$\sim$}}}}

\title{\boldmath Minimal embedding of the Standard Model into intersecting D-brane configurations with a bulk leptonic $U(1)$}

\author[a,b]{I. Antoniadis}
\author[a]{and F. Rondeau}

\affiliation[a]{Laboratoire de Physique Th\'eorique et Hautes Energies - LPTHE\\ Sorbonne Universit\'e, CNRS, 4 Place Jussieu, 75005 Paris, France}

\affiliation[b]{Department of Mathematical Sciences, University of Liverpool,\\ Liverpool L69 7ZL, United Kingdom}

\emailAdd{antoniad@lpthe.jussieu.fr}
\emailAdd{francois.rondeau@lpthe.jussieu.fr}

\abstract{It has been recently shown that the discrepancy between the theoretical and experimental values of the anomalous magnetic moment of the muon can be fully accommodated by considering the contribution of few  Kaluza-Klein (KK) states of the gauged lepton number with masses lighter than the LEP energy, consistently with present experimental limits. In this article, we construct the minimal embedding of the Standard Model (SM) into D-brane configurations with a gauged lepton number. In order to give rise to such KK modes, the lepton number gauge boson must live on an abelian $U(1)_L$ brane extended along at least one ``large'' extra dimension in the bulk, with a compactification scale $M_L\sim\mathcal{O}(10-10^2~{\rm GeV})$ for a string scale $M_s\simgt 10~{\rm TeV}$. As a consequence, $U(1)_L$ cannot participate to the hypercharge linear combination. We show that the minimal realisation of this framework contains five stacks of branes: the SM color $U(3)_c$, weak $U(2)_w$ and abelian $U(1)$ stacks extended effectively only in four dimensions, the bulk $U(1)_L$, as well as a fifth $U(1)^{'}$ brane. With these five abelian factors, one finds besides the hypercharge a second anomaly-free linear combination which does not couple to the SM spectrum, both in the non-supersymmetric case as well as in the minimal supersymmetric extension of the model. It is also shown how the right-handed neutrino can be implemented in the spectrum, and how fermions arising from the two non-SM branes and coupled to the SM through the $U(1)_L$ KK modes can provide Dark Matter candidates. Finally, the possibility of breaking Lepton Flavour Universality is studied by replacing $U(1)_L$ with a brane gauging only the muonic lepton number, avoiding most experimental constraints and enlarging the parameter space for explaining the discrepancy on the muon magnetic moment.
}

\begin{document}
 
\maketitle
\flushbottom

\newpage
\section{Introduction}
\label{sec:intro}

The anomalous magnetic moment of the muon ($(g-2)_{\mu}$) might be one of the most promising signals of new physics beyond the Standard Model (SM). The theoretical calculation, involving QED, electroweak and hadronic contributions, yields a result $a_{\mu}^{{\rm SM}}\equiv (g-2)_{\mu}^{{\rm SM}}/2=116~591~810(43)\times 10^{-11}$ \cite{SM_prediction}, smaller than the experimental measurement from the Brookhaven National Laboratory (BNL) experiment E821 by $3.7\sigma$ \cite{BNL_exp}. The recent Muon $g-2$ experiment at Fermilab has confirmed the BNL results \cite{Fermilab_exp}, pushing the discrepancy with the SM theoretical prediction to $4.7\sigma$, with a difference\footnote{This result must be tempered by the theoretical uncertainties coming from strong interaction effects. The analysis for the hadronic vacuum polarisation contribution has been carried out in \cite{HVP_1,HVP_2,HVP_3}. A recent result from lattice QCD reduce the discrepancy to $1.6 \sigma$ \cite{lattice_QCD}, while producing tensions with other quantities at the same time \cite{lattice_QCD_2,lattice_QCD_3}, so that the hadronic contributions to the muon $(g-2)_{\mu}$ remains an open issue to be addressed by future lattice simulations.} 
\begin{equation}\label{eq:discrepancy}
\Delta a_{\mu}\equiv a_{\mu}^{{\rm exp}}-a_{\mu}^{{\rm SM}}=(2.51\pm 0.59)\times 10^{-9}.
\end{equation}
The possibility to explain this discrepancy in the framework of low mass scale strings and large extra dimensions has been recently studied in \cite{LA_IA_1}, where three contributions to the muon anomalous magnetic moment have been examined : from Regge excitations of the string, from anomalous $U(1)$ gauge bosons, as well as from Kaluza-Klein (KK) modes of a bulk vector field. While the first contribution is strongly suppressed, it has been shown how the second one can reduce, but not fully bridge, the discrepancy. 

However, the most interesting contribution comes from the KK modes of the lepton number gauge boson, denoted $L_{\mu}$ in the following, living on a lepton brane $U(1)_L$ extended along some extra dimensions of the bulk. The zero mode of $L_{\mu}$ is anomalous and acquires a mass through a four-dimensional generalisation of the Green-Schwarz mechanism. If its mass is of order the string scale, it is too heavy to accommodate the $(g-2)_{\mu}$ discrepancy. If it is lighter due to volume suppression, it can only partially explain the discrepancy since the zero mode is a linear combination of the various $U(1)$ factors, coupled to both quarks and leptons and thus subject to stringent LHC bounds~\cite{LA_IA_1, LA_IA_2}. On the other hand, the KK excitations couple only to leptons to lowest order. Therefore, such modes evade the LHC bounds, and their masses can be made sufficiently light to provide a significant contribution to the $(g-2)_{\mu}$ \cite{LA_IA_2}. More precisely, such possibility has been explored in detail in \cite{LA_IA_3}, where it was shown that the discrepancy can be explained assuming the existence of a few leptophilic KK modes lighter than the LEP energy, $\left. \sqrt{s}\right|_{\rm LEP}\sim 200~{\rm GeV}$.

The aim of this paper is to realise the above proposal in the framework of intersecting D-brane models, with matter and gauge fields corresponding to open strings ending on D-branes. More precisely, a single D-brane gives rise to a $U(1)$ gauge theory living on its worldvolume, with the associated gauge boson corresponding to an open string with both ends attached to this brane. Non-abelian gauge symmetries arise from a stack of $N$ coincident D-branes, giving rise to a $U(N)$ gauge symmetry. The matter fields then correspond to open strings stretching between intersecting D-branes, living in their common worldvolume \cite{Inter_Dbrane_1,Inter_Dbrane_2,Inter_Dbrane_3,Inter_Dbrane_4,Inter_Dbrane_5,Inter_Dbrane_6,Inter_Dbrane_7,Inter_Dbrane_8,Inter_Dbrane_9,Inter_Dbrane_10,Inter_Dbrane_11,Inter_Dbrane_12,Inter_Dbrane_13,Inter_Dbrane_14}. In this work, we build the minimal embedding of the Standard Model (SM) into such intersecting D-brane configurations with a gauged lepton number associated to an abelian $U(1)_L$ D-brane. The crucial point to take into account is that such leptonic brane should extend into some extra dimensions transverse to the SM stacks of branes, in order to give rise to the leptophilic KK excitations required in \cite{LA_IA_2,LA_IA_3} to bridge the gap in the muon anomalous magnetic moment. As a consequence, the corresponding $U(1)_L$ cannot contribute to the hypercharge linear combination, since this would lead to an unrealistic small gauge coupling suppressed by the volume of the extra dimensions. Assuming that the lepton number gauge boson propagates into one ``large'' extra dimension, the associated compactification scale $M_L$ must then satisfy $M_L\sim\mathcal{O}(10-10^2~{\rm GeV})$ for a string scale $M_s\simgt 10~{\rm TeV}$ \cite{LA_IA_3}. The total bulk transverse to the SM branes is then made of two inhomogeneous parts. First, the one with one large extra dimension described above along which the $U(1)_L$ brane extends, called ``$L$-bulk'' in the following, whose size $R_L=M_L^{-1}\sim (10-10^2~{\rm GeV})^{-1}$ must be sufficiently low to explain the $(g-2)_{\mu}$ discrepancy. Then, a second part with the remaining (at most five) additional extra dimensions transverse to both SM and $L$ stacks of branes, called ``gravitational bulk'', with an average size larger than the $L$-bulk in order to lower the string scale in the $\simgt\mathcal{O}(10~{\rm TeV})$ region~\cite{ArkaniHamed:1998rs, Antoniadis:1998ig}. From the string theoretic relation $M_{Pl}^2=g_s^{-2}M_s^8V_{(6)}$, with $M_{Pl}$ the four-dimensional Planck mass, $g_s$ the string coupling (assumed to be of order $1$), $M_s$ the string scale and $V_{(6)}=R_L R_G^5$ the volume of the six-dimensional internal space, one gets an average size $R_G$ of the gravitational bulk of the order $R_G\sim (0.1~{\rm GeV})^{-1}$.\footnote{In case of less than six large extra dimensions, $V_{(6)}=R_L R_G^n$ with $n<5$, $R_G$ becomes larger.} In the following, using the term ``bulk'' alone will refer to some extra dimensions transverse to the SM branes when there is no need to specify whether such dimension(s) are in the $L$-bulk or in (part of) the gravitational bulk.

Besides the Standard Model color $U(3)_c$, weak $U(2)_w$ and abelian $U(1)$ stacks of branes localised effectively in four dimensions and the $L$-bulk $U(1)_L$ brane, we show that the minimal embedding of the SM in such configuration requires a fifth $U(1)^{'}$ brane extended in the bulk, so that the total gauge group is $SU(3)_c \times SU(2)_w \times U(1)_c \times U(1)_w \times U(1) \times U(1)^{'} \times U(1)_L$. Identifying $U(1)_c$ with the baryon number $(B)$, we find two possible models, depending on whether the anti-quark $u^C$ or $d^C$ couples to the $U(1)^{'}$. If $U(1)_c$ is not identified with the baryon number, a third model is possible, described here for completeness although not phenomenologically relevant since it forbids the presence of a mass term for the up-type quarks and may lead to dangerous baryon number violating processes. We will thus focus in the models where $U(1)_c$ is identified with the baryon number. In this case, it is shown that there is one anomaly-free $U(1)$ combination besides the hypercharge, which does not couple to the SM spectrum and which can thus remain massless or acquire a mass\footnote{On the other hand, the $U(1)_c$ baryon number is anomalous and acquires a mass through the Green-Schwarz anomaly cancellation mechanism, leaving a global baryon number symmetry preserved at the perturbative level. This mass being of order the string scale, it decouples from the low energy dynamics.}. The minimal supersymmetric extension of our model is also briefly discussed in the context of the anomaly analysis, and we show that the inclusion of the Higgsinos in the spectrum does not modify the result for the non-anomalous vector bosons obtained in the non-supersymmetric case. The presence of two branes extended in the bulk allows to introducing the right-handed neutrino $\nu_R$ as a state of an open string stretched between these two branes. 
When $\nu_R$ is included in the spectrum, the SM particles are now charged under a second anomaly-free $U(1)$ combination given by $B-L$. Depending on the charges of the right-handed neutrino, a third non-anomalous vector can arise, which remains invisible from the SM spectrum as in the situation without $\nu_R$. These two bulk branes can also be used to introduce in a similar way Dark Matter (DM) particle candidates, as Dirac fermions corresponding to open strings stretched between them. By computing the cross-section of the annihilation process of the DM fermions into SM leptons, mediated by the KK excitations of $L_{\mu}$, we find the masses of such DM particles which yield the correct DM relic density, in terms of the compactification scale $M_L$.

As already mentioned above, the anomalous gauge bosons must get a mass through a four-dimensional generalisation of the Green-Schwarz (GS) mechanism~\cite{GS_1,GS_2,GS_3}. Such bosons form in general linear combinations of the various abelian factors associated to each stack of branes, some of them being (effectively) localised in four dimensions while some others propagating into (large) extra dimensions. We will show that such combinations are dominated in the large volume limit by the (zero mode of the) vectors propagating in the bulk which enter in the linear combination defining the anomalous bosons and can become massive with a string scale mass $\mathcal{O}(M_s)$ independent of the compactification scale.

This article is organised as follows. In Section \ref{sect:models}, we construct various D-brane configurations realising the gauge group $SU(3)_c \times SU(2)_w \times U(1)_c \times U(1)_w \times U(1) \times U(1)^{'} \times U(1)_L$, listing the different possibilities for the quantum numbers of the SM spectrum and the allowed Yukawa couplings for the quarks and leptons. The anomaly analysis of these models is performed in Section \ref{sect:anomaly} without the right-handed neutrino, in the non-supersymmetric case as well as in the minimal supersymmetric extension, and then in Section \ref{sect:RH_neutrino} with the inclusion of the right-handed neutrino. Section \ref{sect:mass_spectrum} discusses the gauge bosons mass spectrum arising from the four-dimensional generalisation of the Green-Schwarz mechanism. The inclusion of Dark Matter candidates in this framework is described in Section \ref{sect:DM}. Finally, in Section~\ref{sect:LFNU}, we investigate the possibility of introducing lepton flavour non-universality by gauging only the muonic lepton number $L^{(\mu)}$ that can explain the $(g-2)_{\mu}$ discrepancy due to the exchange of KK excitations that couple only to muons and are thus not constrained by the LEP and LHC bounds. Our conclusions are presented in Section \ref{sect:conclusion}.

\section{The models}
\label{sect:models}
The minimal intersecting D-brane models which can reproduce the SM gauge group $G_{SM}=SU(3)_c\times SU(2)_w\times U(1)_Y$ and its matter spectrum charged under $G_{SM}$ contain three stacks of branes giving rise to a gauge symmetry $U(3)_c\times U(2)_w\times U(1)$ \cite{3_branes_models_1,3_branes_models_2}\footnote{The $U(2)_w$ may be reduced to $Sp(1)\simeq SU(2)$, reducing the number of $U(1)$ factors to two.}. The ``color'' stack $U(3)_c$ and ``weak'' stack $U(2)_w$ are obtained by considering three and two coincident D-branes respectively. For phenomenological reasons, a third $U(1)$ factor arising from a single D-brane is necessary to accommodate the SM.

An open string with one end on the color branes transforms in the $\boldsymbol{3}$ (or $\boldsymbol{\bar 3}$) of $SU(3)_c$; similarly, an open string with one end on the weak branes transforms as a doublet of $SU(2)_w$. The non-abelian structure partially fixes the assignments of the SM particles. The quark doublet $Q$ corresponds to an open string with one end on the color stack and the other on the weak stack of D-branes, while the anti-quark singlets $u^C$ and $d^C$ have one of their ends on the color stack. The lepton doublet $L$ as well as Higgs doublet(s) $H$ must have one of their ends attached to the weak stack of branes. However, the abelian structure is not uniquely determined since the hypercharge can be a linear combination of the different abelian factors.

The standard normalisation for the $U(N)\simeq SU(N)\times U(1)_N$ generators is ${\rm Tr} T^aT^b=\delta^{ab}/2$, while the corresponding $U(1)_N$ charges are measured with respect to the coupling $g_N/\sqrt{2N}$, with $g_N$ the $SU(N)$ coupling constant, so that the fundamental representation of $SU(N)$ has $U(1)_N$ charge unity.

In this work, since we want to obtain leptophilic KK excitations, we require the lepton number gauge boson to be in the bulk and therefore one must consider a bulk leptonic $U(1)_L$ brane. As a starting point, let us consider it to be added to the minimal three stacks model $U(3)_c\times U(2)_w\times U(1)$, namely we consider the total gauge group:
\begin{equation}
G=SU(3)_c \times SU(2)_w \times U(1)_c \times U(1)_w \times U(1) \times U(1)_L.
\end{equation}
Only the lepton doublet $L$ and singlet $e^C$ must have an end attached to the $U(1)_L$. In order to have a lepton number, $L$ and $e^C$ must have opposite $q_L$ charges, choosen to be $+1$ and $-1$ respectively. The other ends of $L$ and $e^C$ must be attached to the $U(2)_w$ and $U(1)$ branes respectively. The $q_1$ charge of $e^C$ and the $q_2\equiv q_w$ charge of $L$ can be defined to be $+1$. The $q_2$ charge $v$ of the quark doublet $Q$ can be either $+1$ or $-1$ if $Q$ belongs to the fundamental $\boldsymbol{2}$ or anti-fundamental $\boldsymbol{\bar 2}$ of $SU(2)_w$. The $q_3\equiv q_c$ charge of $Q$ is fixed to $1$, while the ones of $u^C$ and $d^C$ are chosen to be $-1$ in order to get a baryon number\footnote{The other possible choice $+2$ for the $q_3$ charge of the anti-quarks $u^C$ or $d^C$, which breaks the baryon number symmetry, is discussed below.}. As in the three stacks configuration described above, $u^C$ and $d^C$ must have one of their ends attached on the $U(3)_c$ branes, while the other end can be attached to the $U(1)$ brane ($\{x,y\}=\pm 1$) or be in the bulk ($x=0$ and/or $y=0$). The total matter content and their quantum numbers therefore reads:\footnote{Here we are considering identical embedding for each of the three generations of quarks and leptons.} 
\begin{eqnarray}
&Q& (\boldsymbol{3},\boldsymbol{2};1,v,0,0)_{1/6}\\
&u^C& (\boldsymbol{\bar 3},\boldsymbol{1};-1,0,x,0)_{-2/3}\\
&d^C& (\boldsymbol{\bar 3},\boldsymbol{1};-1,0,y,0)_{1/3}\\
&L& (\boldsymbol{1},\boldsymbol{2};0,1,0,1)_{-1/2}\\
&e^C& (\boldsymbol{1},\boldsymbol{1};0,0,1,-1)_{1},
\end{eqnarray}
with the hypercharge of each species indicated as a subscript for completeness. Since $U(1)_L$ is in the bulk, it should not contribute to the hypercharge, thus given by the combination
\begin{equation}\label{eq:hypercharge}
q_Y=c_3q_3+c_2q_2+c_1q_1,
\end{equation}
where $c_3$, $c_2$ and $c_1$ are constants. The quantum numbers $v$, $x$ and $y$ as well as the constants $c_i$ are now constrained by requiring the different states to have the correct hypercharge.\\

The charges of the leptons $e^C$ and $L$ fixes $c_1=1$ and $c_2=-\frac{1}{2}$ respectively. The charges of the quark doublet $Q$ then imposes $c_3=\frac{1}{6}+\frac{v}{2}$. Finally, the anti-quarks $u^C$ and $d^C$ respectively leads to $x=-\frac{1}{2}+\frac{v}{2}$ and $y=\frac{1}{2}+\frac{v}{2}$. Since $v=\pm 1$, we have either $x=0$ or $y=0$, so that there is at least one end of the $u^C$ or $d^C$ strings elsewhere, which requires the existence of an additional brane $U(1)^{'}$ besides the SM and L branes. To leave open the possibility of having this brane extended in the bulk, in order to avoid again an extremely small gauge coupling, we assume that $U(1)^{'}$ does not participate to the hypercharge, therefore still given by \eqref{eq:hypercharge}. The total gauge group now reads
\begin{equation}
G=SU(3)_c \times SU(2)_w \times U(1)_c \times U(1)_w \times U(1) \times U(1)^{'} \times U(1)_L,
\end{equation}
under which the matter content has the following quantum numbers:
\begin{eqnarray}
&Q& (\boldsymbol{3},\boldsymbol{2};1,v,0,0,0)_{1/6}\\
&u^C& (\boldsymbol{\bar 3},\boldsymbol{1};-1,0,x,z,0)_{-2/3}\\
&d^C& (\boldsymbol{\bar 3},\boldsymbol{1};-1,0,y,w,0)_{1/3}\\
&L& (\boldsymbol{1},\boldsymbol{2};0,1,0,0,1)_{-1/2}\\
&e^C& (\boldsymbol{1},\boldsymbol{1};0,0,1,0,-1)_{1}.
\end{eqnarray}

Again, the constant $v=\pm 1$ specifies whether the quark doublet $Q$ belongs to the fundamental $\boldsymbol{2}$ or anti-fundamental $\boldsymbol{\bar 2}$ representation of $SU(2)_w$. The only ambiguities lie in the quantum numbers of the anti-quarks $u^C$, $d^C$ : they must have one of their ends attached to the $U(3)_c$ branes, while the other ends can be tied to the $U(1)$ or $U(1)^{'}$ branes. This choice, which will be fixed by assigning the correct hypercharges to the states, is encoded in the constants $x,y,z,w=\{0,\pm 1\}$.

We have as previously $c_1=1$, $c_2=-\frac{1}{2}$ and $c_3=\frac{1}{6}+\frac{v}{2}$, while the charges $x$ and $y$ are given by $x=-\frac{1}{2}+\frac{v}{2}$ and $y=\frac{1}{2}+\frac{v}{2}$. Two different models can then be considered, depending on whether $v=+1$ or $v=-1$. 
\begin{itemize}
\item
In the first case $v=+1$, we get $c_3=\frac{2}{3}$, $x=0$ and $y=1$, which enforces $z$ to be non-vanishing and $w=0$. The $u^C$ string has thus one end on the bulk brane $U(1)^{'}$ while $d^C$ is stretched between two branes participating to the hypercharge.
\item
The second case $v=-1$ amounts to exchange the $u^C$ and $d^C$ anti-quarks. Here we get $c_3=-\frac{1}{3}$, $x=-1$ and $y=0$, which implies necessarily that $z=0$ and $w$ is non-vanishing, so that $u^C$ is now stretched between two branes participating to the hypercharge while $d^C$ has one end on the bulk brane $U(1)^{'}$.
\end{itemize}
These two models, defined by the choice $v=\pm 1$, will be respectively denoted A and B in the following. Since $U(1)^{'}$ does not contribute to the hypercharge, the non-vanishing constants $z$ in model A and $w$ in model B can be independently chosen to $\pm 1$. Without lost of generality, we fix them to be $+1$.

\begin{figure}[!h]
    \begin{minipage}[c]{.46\linewidth}
        \centering
        \includegraphics[height=70mm]{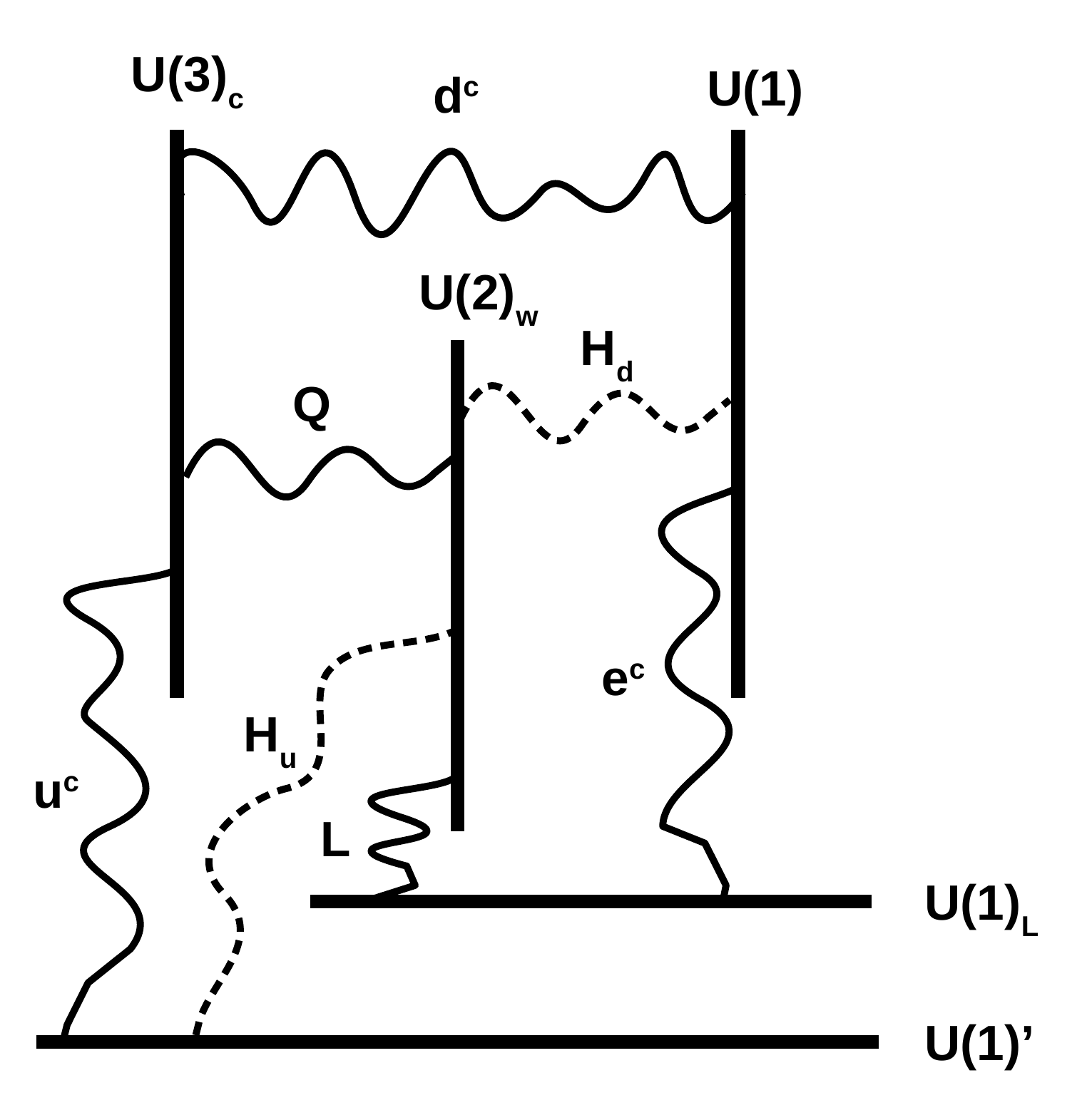}
    \end{minipage}
    \hfill
    \begin{minipage}[c]{.46\linewidth}
        \centering
        \includegraphics[height=70mm]{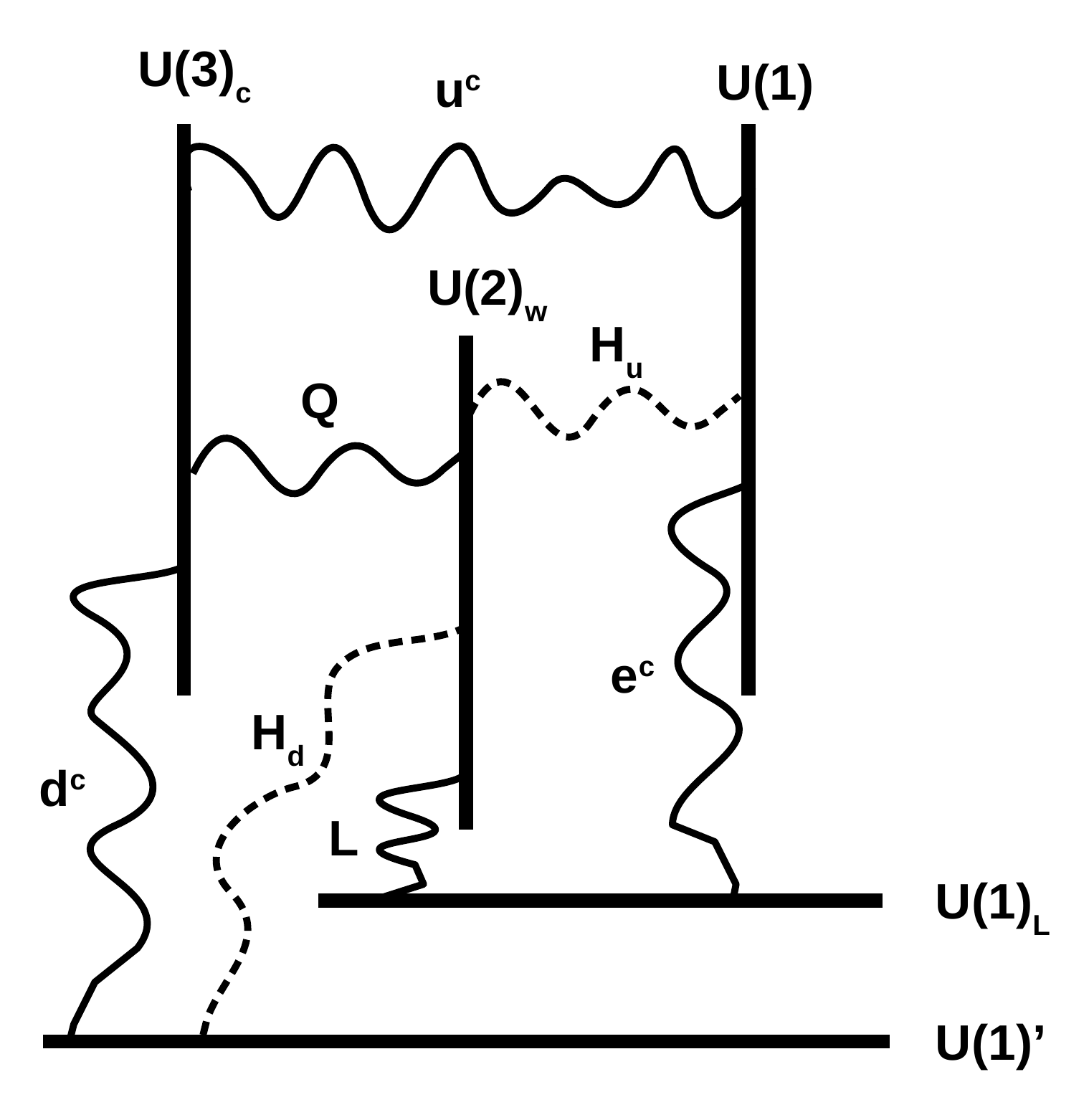}  
    \end{minipage}
    \caption{Pictorial representation of models A and B.}
    \label{fig:pict_rep}
\end{figure}

One can then implement the Higgs doublets. It is easy to check that for each of the models A and B, two Higgs doublets (together with their complex conjugates) with vanishing charges $Q_3$ and $Q_L$ and hypercharge $\pm 1/2$ are possible, given by:
\begin{eqnarray}
\label{eq:Higgs_doublet_model_A}
\text{Model A}&:& H_d (\boldsymbol{1},\boldsymbol{2};0,-1,-1,0,0)_{-1/2},\qquad H_u (\boldsymbol{1},\boldsymbol{2};0,-1,0,-1,0)_{1/2}, \\
\label{eq:Higgs_doublet_model_B}
\text{Model B}&:& H_u (\boldsymbol{1},\boldsymbol{2};0,1,1,0,0)_{1/2},\qquad H_d (\boldsymbol{1},\boldsymbol{2};0,1,0,-1,0)_{-1/2}.
\end{eqnarray}
The allowed Yukawa couplings then read:
\begin{eqnarray}
\label{eq:Yukawa_model_A}
\text{Model A}&:& Qd^CH_d,\qquad Qu^CH_u,\qquad Le^CH_d,\\
\label{eq:Yukawa_model_B}
\text{Model B}&:& Qd^CH_d,\qquad Qu^CH_u,\qquad Le^CH_u^{\dagger}.
\end{eqnarray}
These results are summarized in Tables \ref{tab:model_A} and \ref{tab:model_B}, and the two models are represented pictorially in Figure \ref{fig:pict_rep}.

\begin{center}
\begin{tabular}{|c||ccccc||c|}
   \hline
   $~$ & $q_3$ & \qquad $q_2$ & \qquad $q_1$ & \qquad $q_1'$ & \qquad $q_L$ & $q_Y$ \\
   \hline
   $Q$ & $1$ & \qquad $1$ & \qquad $0$ & \qquad $0$ &  \qquad $0$ & $\frac{1}{6}$ \\
   \hline
   $u^C$ & $-1$ & \qquad $0$ & \qquad $0$ & \qquad $1$ & \qquad $0$ & $-\frac{2}{3}$ \\
   \hline
   $d^C$ & $-1$ & \qquad $0$ & \qquad $1$ & \qquad $0$ & \qquad $0$ & $\frac{1}{3}$ \\
   \hline
   $L$ & $0$ & \qquad $1$ & \qquad $0$ & \qquad $0$ & \qquad $1$ & $-\frac{1}{2}$ \\
   \hline
   $e^C$ & $0$ & \qquad $0$ & \qquad $1$ & \qquad $0$ & \qquad $-1$ & $1$ \\
   \hline
   $H_d$ & $0$ & \qquad $-1$ & \qquad $-1$ & \qquad $0$ & \qquad $0$ & $-\frac{1}{2}$ \\
   \hline
   $H_u$ & $0$ & \qquad $-1$ & \qquad $0$ & \qquad $-1$ & \qquad $0$ & $\frac{1}{2}$ \\
   \hline
\end{tabular}
\captionof{table}{Model A with $q_Y=\frac{2}{3}q_3-\frac{1}{2}q_2+q_1$}
\label{tab:model_A}
\end{center}

\begin{center}
\begin{tabular}{|c||ccccc||c|}
   \hline
   $~$ & $q_3$ & \qquad $q_2$ & \qquad $q_1$ & \qquad $q_1'$ & \qquad $q_L$ & $q_Y$ \\
   \hline
   $Q$ & $1$ & \qquad $-1$ & \qquad $0$ & \qquad $0$ &  \qquad $0$ & $\frac{1}{6}$ \\
   \hline
   $u^C$ & $-1$ & \qquad $0$ & \qquad $-1$ & \qquad $0$ & \qquad $0$ & $-\frac{2}{3}$ \\
   \hline
   $d^C$ & $-1$ & \qquad $0$ & \qquad $0$ & \qquad $1$ & \qquad $0$ & $\frac{1}{3}$ \\
   \hline
   $L$ & $0$ & \qquad $1$ & \qquad $0$ & \qquad $0$ & \qquad $1$ & $-\frac{1}{2}$ \\
   \hline
   $e^C$ & $0$ & \qquad $0$ & \qquad $1$ & \qquad $0$ & \qquad $-1$ & $1$ \\
   \hline
   $H_u$ & $0$ & \qquad $1$ & \qquad $1$ & \qquad $0$ & \qquad $0$ & $\frac{1}{2}$ \\
   \hline
   $H_d$ & $0$ & \qquad $1$ & \qquad $0$ & \qquad $-1$ & \qquad $0$ & $-\frac{1}{2}$ \\
   \hline
\end{tabular}
\captionof{table}{Model B with $q_Y=-\frac{1}{3}q_3-\frac{1}{2}q_2+q_1$}
\label{tab:model_B}
\end{center}

\begin{figure}[!h]
        \centering
        \includegraphics[height=70mm]{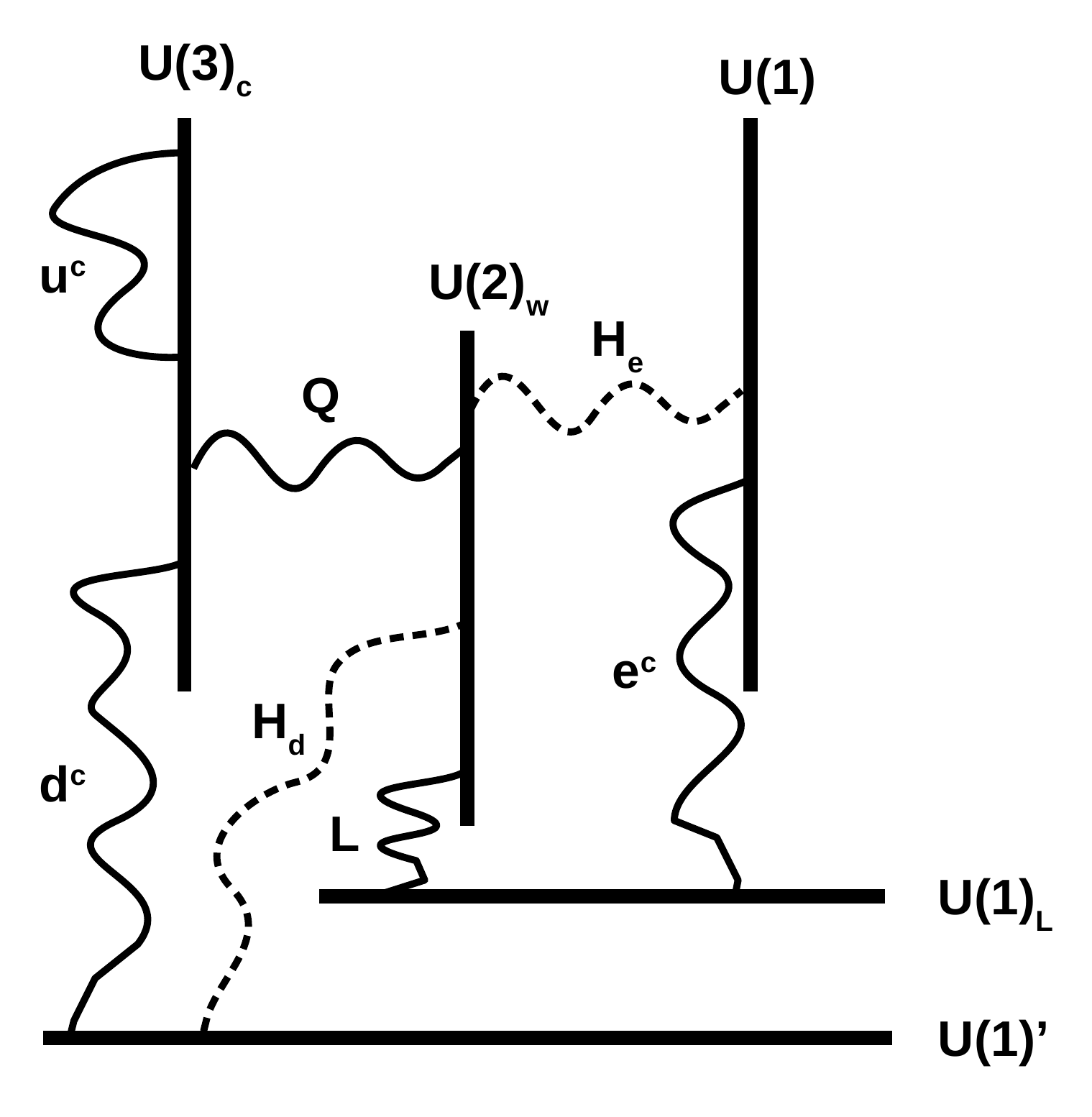}
    \caption{Pictorial representation of model C.}
    \label{fig:pict_rep_C}
\end{figure}

One can also consider a model where the anti-quark $u^C$ has both ends attached to the $U(3)$ stack of branes and its orientifold image, corresponding to the $\boldsymbol{\bar 3}$ of $SU(3)$ obtained as the antisymmetric product of two $\boldsymbol{3}$'s. Repeating the analysis described above, we get the matter content summarized in Table \ref{tab:model_C}.
\begin{center}
\begin{tabular}{|c||ccccc||c|}
   \hline
   $~$ & $q_3$ & \qquad $q_2$ & \qquad $q_1$ & \qquad $q_1'$ & \qquad $q_L$ & $q_Y$ \\
   \hline
   $Q$ & $1$ & \qquad $-1$ & \qquad $0$ & \qquad $0$ &  \qquad $0$ & $\frac{1}{6}$ \\
   \hline
   $u^C$ & $2$ & \qquad $0$ & \qquad $0$ & \qquad $0$ & \qquad $0$ & $-\frac{2}{3}$ \\
   \hline
   $d^C$ & $-1$ & \qquad $0$ & \qquad $0$ & \qquad $1$ & \qquad $0$ & $\frac{1}{3}$ \\
   \hline
   $L$ & $0$ & \qquad $1$ & \qquad $0$ & \qquad $0$ & \qquad $1$ & $-\frac{1}{2}$ \\
   \hline
   $e^C$ & $0$ & \qquad $0$ & \qquad $1$ & \qquad $0$ & \qquad $-1$ & $1$ \\
   \hline
   $H_e$ & $0$ & \qquad $1$ & \qquad $1$ & \qquad $0$ & \qquad $0$ & $\frac{1}{2}$ \\
   \hline
   $H_d$ & $0$ & \qquad $-1$ & \qquad $0$ & \qquad $1$ & \qquad $0$ & $\frac{1}{2}$ \\
   \hline
\end{tabular}
\captionof{table}{Model C with $q_Y=-\frac{1}{3}q_3-\frac{1}{2}q_2+q_1$}
\label{tab:model_C}
\end{center}
In this case, an up quark mass term is no longer allowed since $Qu^C$ has a non-vanishing $q_3$ charge. The only possible Yukawa couplings are
\begin{equation}
Qd^CH_d^{\dagger},\qquad Le^CH_e^{\dagger},
\end{equation}
where $H_d$ and $H_e$ have been defined in Table \ref{tab:model_C}. Let us note that in this model, the role of $u^C$ and $d^C$ is not symmetric, and there is no consistent model with $d^C$ in the $\boldsymbol{\bar 3}$ of $SU(3)$. The model C is represented pictorially in Figure \ref{fig:pict_rep_C}.

\section{Anomaly analysis}
\label{sect:anomaly}
In order to find the anomalous and anomaly free combinations of the $U(1)$'s in the different models constructed above, we compute the anomaly matrix $K_{IJ}={\rm Tr} T_I^2Q_J$, built from the mixed gauge and gravitational anomalies. The column of indices $J=\left\{c,w,1,1^{'},L\right\}$ labels the abelian generators $Q_J$, while the line indices are $I=\left\{SU(3), SU(2), Y,~{\rm Grav}\right\}$, with $T_{{\rm Grav}}=\mathbb{1}$ for the gravitational anomalies. One gets, for the two models A and B:
\begin{eqnarray}
\text{Model A}&:& 
K^{(A)} = \begin{pmatrix}
0 & 1 & \frac{1}{2} & \frac{1}{2} & 0 \\
\frac{3}{2} & 2 & 0 & 0 & \frac{1}{2} \\
-\frac{3}{2} & \frac{2}{3} & \frac{4}{3} & \frac{4}{3} & -\frac{1}{2} \\
0 & 8 & 4 & 3 & 1 
\end{pmatrix}, \\
\text{Model B}&:& 
K^{(B)} = \begin{pmatrix}
0 & -1 & -\frac{1}{2} & \frac{1}{2} & 0 \\
\frac{3}{2} & -1 & 0 & 0 & \frac{1}{2} \\
-\frac{3}{2} & \frac{1}{3} & -\frac{1}{3} & \frac{1}{3} & -\frac{1}{2} \\
0 & -4 & -2 & 3 & 1 
\end{pmatrix}.
\end{eqnarray}
Diagonalizing the matrices $K^{\top}K$, one finds the anomaly free $U(1)$'s as the eigenvectors associated to the vanishing eigenvalues, the other eigenvectors being anomalous. In addition to the hypercharge $q_Y$, one finds in both models a second anomaly free $U(1)$ given by:
\begin{eqnarray}
\label{eq:anomaly_free_without_nu_R}
\text{Model A}&:& q\equiv\frac{1}{3}q_3-\frac{1}{2}q_2+q_1^{'}+q_L,\\
\label{eq:anomaly_free_without_nu_R_model_B}
\text{Model B}&:& q\equiv-\frac{2}{3}q_3-\frac{1}{2}q_2-q_1^{'}+q_L.
\end{eqnarray}
It is easy to see that the $q$ charges of the SM particles are minus their hypercharge $q_Y$, namely the SM sees only one anomaly free $U(1)$ as expected. The second anomaly free combination reads $q_3-q_2+q_1+q_1^{'}+q_L$, and is invisible from the SM as one can easily check.\\

The analysis in the model C is carried out in a similar way. In that case, the anomaly matrix reads
\begin{equation}
K^{(C)}=
\begin{pmatrix}
\frac{3}{2} & -1 & 0 & \frac{1}{2} & 0 \\
\frac{3}{2} & -1 & 0 & 0 & \frac{1}{2} \\
\frac{5}{2} & \frac{1}{3} & 1 & \frac{1}{3} & -\frac{1}{2} \\
9 & -4 & 1 & 3 & 1 
\end{pmatrix},
\end{equation}
and it turns out that the hypercharge $q_Y$ is the only anomaly free $U(1)$.

One can wonder whether the supersymmetrisation of our models modify the results of the anomaly analysis. In the Minimal Supersymmetric extension of the Standard Model (MSSM), each SM particle gets a superpartner : the quark and lepton fermions are paired with the sleptons and squarks, the SM gauge bosons are paired with the gauginos, while the spin-$1/2$ fermionic partners of the Higgs scalars are the Higgsinos. Being chiral, the Higgsinos $\tilde H$ can modify the anomalous and anomaly-free $U(1)$ combinations obtained in the non-supersymmetric models. In our case, their introduction in the spectrum gives new contributions to the entries $K_{IJ}$ for $I,J=\{2,3,4\}$ of the anomaly matrix $K$. For model A, where the quantum numbers of the Higgsinos are
\begin{equation}\label{eq:Higgsinos_model_A}
\tilde H_d (\boldsymbol{1},\boldsymbol{2};0,-1,-1,0,0)_{-1/2},\qquad \tilde H_u (\boldsymbol{1},\boldsymbol{2};0,-1,0,-1,0)_{1/2},
\end{equation}
one gets the anomaly matrix
\begin{equation}
K^{(A)}_{{\rm MSSM}} = \begin{pmatrix}
0 & 1 & \frac{1}{2} & \frac{1}{2} & 0 \\
\frac{3}{2} & 1 & -\frac{1}{2} & -\frac{1}{2} & \frac{1}{2} \\
-\frac{3}{2} & -\frac{1}{3} & \frac{5}{6} & \frac{5}{6} & -\frac{1}{2} \\
0 & 4 & 2 & 1 & 1 
\end{pmatrix}.
\end{equation} 
It is then easy to check that the vectors associated to the zero eigenvalues of the matrix $\left(K^{(A)}_{{\rm MSSM}}\right)^{\top}K^{(A)}_{{\rm MSSM}}$ are $q_y$ and $q$ given by \eqref{eq:anomaly_free_without_nu_R}, so that the minimal supersymmetric extension of model A does not modify the result obtained in the non-supersymmetric case.

On the other hand, the minimal supersymmetric extension of model B would require the introduction of a third Higgs doublet, since with the two Higgs doublets \eqref{eq:Higgs_doublet_model_B} the Yukawa couplings \eqref{eq:Yukawa_model_B} would violate holomorphy of the superpotential.

Let us note that an anomaly-free $U(1)$ is not necessarily massless because of underlying $6$-dimensional anomalies \cite{6d_anomalies_1,6d_anomalies_2,6d_anomalies_3}. One therefore needs to impose additional model-dependent constraints to ensure that the hypercharge remains massless in four dimensions. As an example, let us assume that our framework arises from a given type IIA orientifold compactification 
with D6-branes and orientifold O6-planes: we denote by $i$ a stack of $N_i$ D6-branes giving rise to a factor $U(N_i)$ in the gauge group, and $\tilde i$ its orientifold image. The D6$_i$-branes span 4-dimensional Minkowski space and are wrapped on 3-cycles $\Pi_i$ in the internal space $X_6$. In general, 3-cycles in 6-dimensional compact space intersect several times. Introducing the 3-homology class $[\Pi_i]$ of the 3-cycle $\Pi_i$, the homological intersection number of the stacks $i$ and $j$ is given by $I_{ij}=[\Pi_i]\cdot[\Pi_j]$. The data $N_i$ and $I_{ij}$ are then sufficient to determine the massless chiral spectrum of the 4-dimensional theory
\footnote{We do not discuss the sector $i~\tilde i$ which does not play a role in our analysis.}:
\ytableausetup{smalltableaux}
\begin{itemize}
\item $ij$ sector: open strings stretching between the stacks $i$ et $j$ correspond to $I_{ij}$ $4D$ chiral fermions in the bi-fundamental representation $(\bold{N_i},\bold{\overline{N}_j})$ of $U(N_i)\times U(N_j)$.
\vspace{0.2cm}
\item $i~\tilde{j}$ sector: open strings stretching between the stacks $i$ and the orientifold image $\tilde j$ of the stack $j$ correspond to $I_{i\tilde j}$ $4D$ chiral fermions in the bi-fundamental representation $(\bold{N_i},\bold{N_j})$ of $U(N_i)\times U(N_j)$.
\end{itemize} 
In this framework, 
an anomaly free $U(1)$ linear combination
\begin{equation}\label{}
q_Y=\sum_i c_i q_i
\end{equation}
remains massless if the following condition holds \cite{massless_U(1)_constraint}:
\begin{equation}\label{eq:massless_U(1)_constraint}
\sum_{i\neq j} c_i N_i \left(I_{ji}-I_{j\tilde i}\right)=0\,,
\end{equation}
for every $j$, where the sum runs over $i$. 
Let us take as an example the minimal supersymmetric extension of model B built in Section~\ref{sect:models}, where the constants $c_i$ defining the hypercharge linear combination are $c_3=-\frac{1}{3}$, $c_2=-\frac{1}{2}$, $c_1=1$ and $c_{1'}=c_{1_L}=0$. Using the intersection numbers corresponding to the fermionic spectrum of Table \ref{tab:model_B},
\begin{equation}\label{eq:intersection_numbers}
I_{32}=3,\quad I_{3\tilde 1}=-3,\quad I_{31'}=-3, \quad I_{2\tilde{1}_L}=3, \quad I_{11_L}=3,
\end{equation}
one can easily check that the constraints \eqref{eq:massless_U(1)_constraint} for $j=\{3,1_L\}$ are indeed satisfied. With the intersection numbers $I_{2\tilde{1}}=1$ and $I_{21'}=1$, corresponding to the two Higgsinos doublet superpartners of the Higgs scalars $H_u$ and $H_d$ defined in Table \ref{tab:model_B}, the constraints \eqref{eq:massless_U(1)_constraint} for $j=\{2,1,1'\}$ are not satisfied. It would be the case if we could have for instance the intersection number $I_{2\tilde{1}}=3$ and $I_{21'}=3$. This amounts of introducing two additional Higgs doublet pairs in the $(\bold{2},\bold{1})$ and $(\bold{2},\bold{1'})$, which do not modify our phenomenological analysis presented here.

Note finally that the local D-brane configurations built in this paper may also be obtained from other string constructions which do not admit such interpretation in terms of intersecting D-branes (such as ordinary type I orbifolds, or non-commuting magnetized D-branes), for which the above conditions do not apply as such.

\section{Right-handed neutrino}
\label{sect:RH_neutrino}
Finally, we can implement the right-handed neutrino $\nu_R$. The existence in the total gauge group of two abelian factors which do not participate to the hypercharge easily allows to introduce such SM singlet state $\nu_R$, corresponding either to an open string with one end on the $U(1)^{'}$ brane and the other on the $U(1)_L$ brane, or with both ends on one of the two branes.\footnote{For notational simplicity, we call a charged open string with ends on the same brane when it stretches between the brane and its orientifold image.} The right-handed neutrino enters the anomaly analysis only through the gravitational anomalies, modifying the entries $K_{44}$ and $K_{45}$ of the anomaly matrices computed above. In the following we will focus on model A, the analysis in the two other models can be carried out in a completely similar way and does not bring any new relevant physical results. In the case when the open string associated to $\nu_R$ stretches between the $U(1)_L$ and $U(1)^{'}$ branes, the four different possibilities for the charge assignments of $\nu_R$ together with the associated anomaly-free $U(1)$ combinations is listed in Table \ref{tab:nu_R_config}\footnote{It is easy to check that these results remain unchanged in the minimal supersymmetric extension of the model, introducing the Higgsinos \eqref{eq:Higgsinos_model_A} in the spectrum.}.
\begin{center}
\begin{tabular}{|c||c||c|}
   \hline
   Configurations & $\nu_R$ quantum numbers & Anomaly-free $U(1)$ (besides the hypercharge) \\
   \hline
   1 & $\nu_R (\boldsymbol{1},\boldsymbol{1};0,0,0,-1,1)_{0}$ & $q\equiv\frac{1}{3}q_3-\frac{1}{2}q_2+q_1^{'}+q_L$ \\
   \hline
   2 & $\nu_R (\boldsymbol{1},\boldsymbol{1};0,0,0,1,1)_{0}$ & $\tilde q\equiv\frac{2}{3}q_3-\frac{1}{2}q_2+q_1^{'}$  \\
   \hline
   3 & $\nu_R (\boldsymbol{1},\boldsymbol{1};0,0,0,-1,-1)_{0}$ & $-\frac{1}{3}q_3+q_L$ \\
   \hline
   4 & $\nu_R (\boldsymbol{1},\boldsymbol{1};0,0,0,1,-1)_{0}$ & $\tilde q\equiv\frac{2}{3}q_3-\frac{1}{2}q_2+q_1^{'}$ \quad ; \quad $-\frac{1}{3}q_3+q_L$ \\
   \hline
\end{tabular}
\captionof{table}{Anomaly-free $U(1)$ in configurations with right-handed neutrinos (model A)}
\label{tab:nu_R_config}
\end{center}
The first configuration does not modify the result \eqref{eq:anomaly_free_without_nu_R} obtained in the absence of $\nu_R$ : the SM particles see one anomaly-free $U(1)$, the hypercharge $q_Y$, while there is a second anomaly free combination $q_3-q_2+q_1+q_1^{'}+q_L$ invisible from the SM. The second configuration contains also an extra anomaly-free $U(1)$, $\tilde q\equiv\frac{2}{3}q_3-\frac{1}{2}q_2+q_1^{'}$. In that case, one observes that the $\tilde q$ charges of the SM particles are given by $B-L-q_Y$\footnote{The $B-L$ charges are defined as usual : $1/3$ for $Q$, $-1/3$ for $u^C$ and $d^C$, $-1$ for $L$ and $1$ for $e^C$ and $\nu_R$.}: the SM spectrum is thus charged under two anomaly-free $U(1)$'s, $q_Y$ and $B-L$. The situation is similar in the third configuration, with $B-L$ still anomaly free and now given by $\frac{1}{3}q_3-q_L$. Finally, the fourth configuration combines the features of the three previous ones : there are now two anomaly-free $U(1)$'s besides the hypercharge, $\tilde q\equiv\frac{2}{3}q_3-\frac{1}{2}q_2+q_1^{'}$ and $\frac{1}{3}q_3-q_L$. The $\tilde q$ charges of the SM particles are given, as in the second configuration, by $B-L-q_Y$, so that the SM sees again two anomaly-free $U(1)$'s, $q_Y$ and $B-L$. The third anomaly-free combination, invisible from the SM, is given as in the first configuration by $q_3-q_2+q_1+q_1^{'}+q_L$. 

In configurations 2, 3 and 4, the anomaly-free $B-L$ boson may then acquire a mass as a consequence of 6-dimensional anomalies as mentioned at the end of Section \ref{sect:anomaly}, or through a standard Higgs mechanism\footnote{In this case its mass can be much lower than the string scale and $B-L$ should be included in the low energy theory and could mix with the hypercharge. Standard Model extensions with such $U(1)$'s have been extensively studied in the literature and their analysis goes beyond the scope of our paper.}.

Regarding the $\nu_R$ mass, only the fourth configuration allows for a Yukawa coupling $L\nu^RH_u$. One way to obtain a small Dirac neutrino mass compatible with the experimental bounds is to allow $\nu^R$ to propagate in the bulk, in which case the Dirac mass $m_{\nu_R}$ coming from such Yukawa coupling is suppressed by the volume $V_{\perp}$ of the extra transverse dimensions, namely $m_{\nu_R}\sim\frac{v}{\sqrt{V_{\perp}}}$, with $v$ the vev of the Higgs field \cite{neutrino_masses}. This can be obtained in our framework if the $U(1)^{'}$ brane extends along the extra dimension of the $U(1)_L$, so that $\nu_R$ propagates in the $L$-bulk. In that case however, the $L$-bulk having one extra dimension, it is easy to see that such Dirac neutrino mass is much above the upper limit $\sum m_{\nu_i}\simlt 0.1~{\rm eV}$, so that the fourth configuration is phenomenologically excluded. A Yukawa coupling $L\nu^RH_u$ is forbidden in the three first cases, since such a term would not be neutral under $U(1)^{'}$ or $U(1)_L$. However, a Dirac mass term can still arise through non-perturbative effects, taking the form $L\nu^RH_u e^{-\frac{\alpha}{g_s}}$, with $g_s$ the string coupling and $\alpha$ a model-dependent numerical factor \cite{instanton_dirac_neutrino_masses}.

On the other hand, neutrino Majorana masses are perturbatively forbidden since such terms break the (global) lepton number symmetry, but can also arise from non-perturbative instanton effects \cite{instanton_majorana_neutrino_masses}.
 
\section{Mass spectrum}
\label{sect:mass_spectrum}
The $U(1)$ combinations orthogonal to the anomaly free $U(1)$'s, $Y$, \eqref{eq:anomaly_free_without_nu_R}, \eqref{eq:anomaly_free_without_nu_R_model_B} or listed in Table \ref{tab:nu_R_config}, are anomalous and acquire a mass through a 4-dimensional generalisation of the Green-Schwarz (GS) mechanism. In our model, such anomalous bosons form linear combination of some $U(1)$ localised in 4-dimensions and some others extended in the bulk. The aim of this section is to clarify how this situation impacts the anomaly analysis, and in particular if the compactification scale enters the mass of the anomalous bosons.

To simplify the analysis, let us consider a toy model with three $U(1)$ bosons : $B_{\mu}(x)$ and $C_{\mu}(x)$ are localised in four dimensions, while $X_{\mu}(x,y)$ is a bulk vector. Their $U(1)$ charges are respectively denoted $Q_3$, $Q_2$ and $Q_1$, while their kinetic action is given by
\begin{eqnarray}\label{eq:kin_term}
S_{kin}=\int d^4x \left[-\frac{1}{4g_3^2}F^2(B)-\frac{1}{4g_2^2}F^2(C)\right]-\frac{1}{4g_{1(5)}^2}\int d^5x F^2(X).
\end{eqnarray}
The standard KK reduction is carried out by expanding $X_{\mu}$ and $X_5$ according to $X_{\mu}(x,y)=\sum_{n\in\mathbb{Z}}X_{\mu}^{(n)}(x)e^{\frac{iny}{R}}$ and $X_{5}(x,y)=\sum_{n\in\mathbb{Z}}X_{5}^{(n)}(x)e^{\frac{iny}{R}}$. Integrating then the second term of \eqref{eq:kin_term} over $y$, we get:
\begin{eqnarray}
S_{kin}&=&\int d^4x \left[-\frac{1}{4g_3^2}F^2(B)-\frac{1}{4g_2^2}F^2(C)\right]\nonumber\\
&&\quad-\frac{1}{4g_1^2}\int d^4x \left[F^2_{(0)}(X)+\sum_{n\neq 0}F^{(n)}_{\mu\nu}F^{\mu\nu(-n)}+2\left(\partial_{\mu}X_{5}^{(0)}\right)^2\right]\\
&&\qquad-\frac{1}{2g_1^2}\int d^4x \sum_{n\neq 0}\left(\partial_{\mu}X_5^{(n)}(x)-\frac{in}{R}X_{\mu}^{(n)}(x)\right)\left(\partial_{\mu}X_5^{(-n)}(x)+\frac{in}{R}X_{\mu}^{(-n)}(x)\right)\nonumber,
\end{eqnarray}
where we have defined the four-dimensional effective gauge coupling $g_1$ from the five-dimensional one $g_{1(5)}$ by
\begin{equation}\label{eq:effective_coupling}
\frac{1}{g_1^2}=\frac{V_{\perp}}{g_{1(5)}^2},
\end{equation}
with $V_{\perp}$ the volume of the extra dimensions. In the gauge in which $\forall n\neq 0$, $X_5^{(n)}=0$, the KK excitations $X_{\mu}^{(n)}$, $n\neq 0$, become massive. The gauge symmetries associated to these states having been fixed, the bosons $X_{\mu}^{(n)}$, $n\neq 0$, do not contribute to anomalies, so that only combinations of $B_{\mu}(x)$, $C_{\mu}(x)$ and the zero mode $X_{\mu}^{(0)}(x)$ of $X_{\mu}(x,y)$ can be anomalous. In the following, we will denote the basis $(B_{\mu}, C_{\mu}, X_{\mu}^{(0)})$ as the ``D-brane basis''.

We next consider the basis formed by the hypercharge $Y_{\mu}$ and two anomalous vectors $A_{\mu}$ and $A_{\mu}^{'}$, all of three orthogonal to each other, denoted ``hypercharge basis'' in the following. In the most general case, the hypercharge $Y_{\mu}$ is a linear combination of all the bosons of the D-brane basis localised in four dimensions, namely in this model $B_{\mu}(x)$ and $C_{\mu}(x)$, while $A_{\mu}$ and $A_{\mu}^{'}$ can be combinations of all the D-brane basis bosons, including $X_{\mu}^{(0)}(x)$. We parametrise these combinations as:
\begin{subequations}\label{eq:charge_relations}
\begin{align}
Q_Y&=c_3Q_3+c_2Q_2,\\
Q_A&=c_2Q_3-c_3Q_2+c_1Q_1,\\
Q_{A^{'}}&=c_2Q_3-c_3Q_2-\frac{c_2^2+c_3^2}{c_1}Q_1.
\end{align}
\end{subequations}
In order to relate the original D-brane basis $(B_{\mu}, C_{\mu}, X_{\mu}^{(0)})$ to the hypercharge basis \allowbreak $(Y_{\mu}, A_{\mu}, A^{'}_{\mu})$, we write the covariant derivatives of the bosons in both bases (assuming a canonical normalisation of their kinetic terms): 
\begin{eqnarray}
\mathcal{D}_{\mu}&=&\partial_{\mu}-i\frac{g_3}{\sqrt{6}}Q_3B_{\mu}(x)-i\frac{g_2}{2}Q_2C_{\mu}(x)-ig_1Q_1X_{\mu}^{(0)}(x)\\
&=&\partial_{\mu}-ig_YQ_YY_{\mu}(x)-ig_AQ_AA_{\mu}(x)-ig_{A^{'}}Q_{A^{'}}A^{'}_{\mu}(x).
\end{eqnarray}
Using the relations \eqref{eq:charge_relations}, one can identify the different terms and get the resulting $3\times 3$ rotation matrix $\mathcal{R}$ relating $(B_{\mu}, C_{\mu}, X_{\mu}^{(0)})$ to $(Y_{\mu}, A_{\mu}, A^{'}_{\mu})$:
\begin{equation}\label{eq:rotation_matrix}
\begin{pmatrix}
Y_{\mu} \\
A_{\mu} \\
A^{'}_{\mu}
\end{pmatrix}=
\begin{pmatrix}
\sqrt{6}c_3\frac{g_Y}{g_3} & 2c_2\frac{g_Y}{g_2} & 0 \\
\sqrt{6}c_2\frac{g_A}{g_3} & -2c_3\frac{g_A}{g_2} & c_1\frac{g_A}{g_1} \\
\sqrt{6}c_2\frac{g_{A^{'}}}{g_3} & -2c_3\frac{g_{A^{'}}}{g_2} & -\frac{c_2^2+c_3^2}{c_1}\frac{g_{A^{'}}}{g_1}
\end{pmatrix}
\begin{pmatrix}
B_{\mu} \\
C_{\mu} \\
X_{\mu}^{(0)}
\end{pmatrix}.
\end{equation}
Imposing the orthogonality condition for $\mathcal{R}$, $\sum_{j=1}^3\mathcal{R}_{ij}^2=1, \forall i=1,2,3$, we get the well-known relations between the coupling constants of the vector bosons in the two bases :
\begin{subequations}
\begin{align}
\frac{1}{g_Y^2}&= \frac{6c_3^2}{g_3^2}+\frac{4c_2^2}{g_2^2},\\
\frac{1}{g_A^2}&= \frac{6c_2^2}{g_3^2}+\frac{4c_3^2}{g_2^2}+\frac{c_1^2}{g_1^2},\\
\frac{1}{g_{A^{'}}^2}&= \frac{6c_2^2}{g_3^2}+\frac{4c_3^2}{g_2^2}+\left(\frac{c_2^2+c_3^2}{c_1}\right)^2\frac{1}{g_1^2}.
\end{align}
\end{subequations}
Using then \eqref{eq:effective_coupling} relating the five dimensional coupling constant $g_{1(5)}$ and the four dimensional one $g_1$, one sees that the relations for $g_A$ and $g_{A^{'}}$ are dominated in the large volume limit by the $g_1$ coupling, namely
\begin{subequations}\label{eq:coupl_const_large_vol_lim}
\begin{align}
\frac{1}{g_Y^2}&= \frac{6c_3^2}{g_3^2}+\frac{4c_2^2}{g_2^2},\\
\frac{1}{g_A^2}&\sim \frac{c_1^2}{g_1^2}=V_{\perp}\frac{c_1^2}{g_{1(5)}^2},\\
\frac{1}{g_{A^{'}}^2}&\sim \left(\frac{c_2^2+c_3^2}{c_1}\right)^2\frac{1}{g_1^2}=V_{\perp}\left(\frac{c_2^2+c_3^2}{c_1}\right)^2\frac{1}{g_{1(5)}^2}.
\end{align}
\end{subequations}
The above rotation matrix in Eq. \eqref{eq:rotation_matrix} has thus the following structure:
\begin{equation}
\begin{pmatrix}
Y_{\mu} \\
A_{\mu} \\
A^{'}_{\mu}
\end{pmatrix}\sim
\begin{pmatrix}
\mathcal{O}(1) & \mathcal{O}(1) & 0 \\
\mathcal{O}\left(\frac{1}{\sqrt{V_{\perp}}}\right) & \mathcal{O}\left(\frac{1}{\sqrt{V_{\perp}}}\right) & \mathcal{O}(1) \\
\mathcal{O}\left(\frac{1}{\sqrt{V_{\perp}}}\right) & \mathcal{O}\left(\frac{1}{\sqrt{V_{\perp}}}\right) & \mathcal{O}(1)
\end{pmatrix}
\begin{pmatrix}
B_{\mu} \\
C_{\mu} \\
X_{\mu}^{(0)}
\end{pmatrix},
\end{equation}
so that, in the large volume limit, the bosons $A_{\mu}$ and $A_{\mu}^{'}$ are simply given by the zero mode of the bulk vector $X_{\mu}$.\\

The anomalies being localised in four dimensions, the effective action involved in the GS anomaly cancellation is given by
\begin{eqnarray}
S&=&\int d^4x \left[-\frac{1}{4g_A^2}F_A^2-\frac{1}{2}\left(\partial_{\mu}a+M_sA_{\mu}\right)^2+\frac{a}{M_s}\sum_{I}k_I {\rm Tr}F_I\wedge F_I\right]\nonumber\\
&&\quad +\int d^4x \left[-\frac{1}{4g_{A^{'}}^2}F_{A^{'}}^2-\frac{1}{2}\left(\partial_{\mu}a^{'}+M_sA_{\mu}^{'}\right)^2+\frac{a^{'}}{M_s}\sum_{I}k_I^{'} {\rm Tr}F_I\wedge F_I\right],
\end{eqnarray}
where $F_A$ ($F_{A^{'}}$) is the field strength of the anomalous $U(1)_A$ ($U(1)_{A^{'}}$), $g_A$ ($g_{A^{'}}$) the associated gauge coupling, and $a$ ($a^{'}$) the pseudo-scalar axion responsible for the anomaly cancellation. The indice $I$ in the sum over Pontryagin densities denotes $SU(3)$, $SU(2)$ and $Y$ for the mixed gauge anomalies, $F_I$ are the associated field strengths and the constants $k_I$ ($k_I^{'}$) are given by $k_I={\rm Tr}T_I^2Q_A$ ($k_I^{'}={\rm Tr}T_I^2Q_{A^{'}}$)\footnote{The gravitational anomalies are also canceled by a similar term of the form $\frac{a}{M_s}k_G R\wedge R$, where $k_G={\rm Tr}Q_A$, and similarly for $A^{'}$, $a^{'}$.}.

In the most general case where both the axion and the anomalous vector propagate into some extra dimensions of the bulk, the mass $M_A$ of the anomalous gauge boson is of order $M_A\propto \sqrt{\frac{V_a}{V_A}}M_s$, where $V_a$ and $V_A$ denote the volume of the space where the axion $a$ and the vector $A_{\mu}$ propagate, respectively \cite{6d_anomalies_1,Anastasopoulos:2004ga,Anastasopoulos:2006cz}. If $a$ is localised in four dimensions, then $M_A\propto \frac{M_s}{\sqrt{V_A}}$, and thus, for one extra dimension at a scale $M_L \sim 10~{\rm GeV}$ and a string scale $M_s\sim 10~{\rm TeV}$, we get $M_A\sim 10^2~{\rm GeV}$. This mass being too low and subject to stringent phenomenological constraints, one needs to have $V_a=V_A$ in order to get a mass of the anomalous boson of the order of the string scale. We conclude that the axions which cancel the anomalies of anomalous $U(1)$ combinations which have a component along $L_{\mu}$ must also propagate in the $L$-bulk.

\section{Dark Matter model}
\label{sect:DM}

\begin{figure}[!h]
        \centering
        \includegraphics[height=35mm]{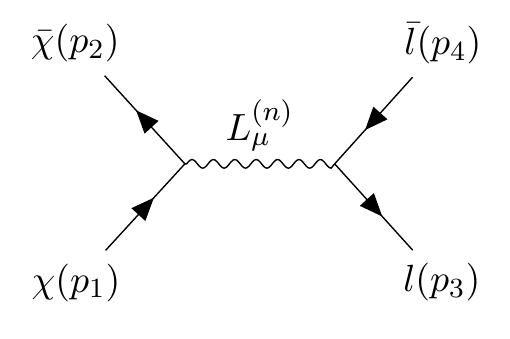}
    \caption{Dark matter annihilation into SM leptons mediated by the KK excitations $L_{\mu}^{(n)}$ of the lepton number gauge boson.}
    \label{fig:feynman_diag}
\end{figure}

In this section, we briefly describe how the models built in this paper can easily provide Dark Matter (DM) candidates, as Dirac fermions (called $\chi$ in the following) corresponding to open strings stretched between bulk branes, similar to the right-handed neutrinos. The simplest situation arises when $\chi$ has one end on the $U(1)_L$ brane and the other on the $U(1)^{'}$ brane. If the extra dimension(s) along which the $U(1)^{'}$ extends is (are) orthogonal to the $L$-bulk, then $\chi$ is localised in four dimensions and has no KK modes. The KK excitations $L_{\mu}^{(n)}$ of the lepton number gauge boson, which couple to both $\chi$ and the SM leptons $l$, then provide a mediator between the dark and the (leptonic) visible sector, allowing DM annihilation process $\chi\bar{\chi}\rightarrow l\bar l$ through the s-channel diagram represented in Figure \ref{fig:feynman_diag}. Its amplitude is given by:
\begin{equation}
\mathcal{M}_n=\frac{-i}{s-n^2M_L^2}\left[\bar{v}(p_2)(-ig_L\gamma^{\mu})u(p_1)\right]\left[\bar{u}(p_3)(-ig_L\gamma_{\mu})v(p_4)\right],
\end{equation}
with 
$M_L$ the compactification scale of the extra dimension of the $L$-bulk where $L_{\mu}$ propagates, $u$ and $v$ the on-shell external spinors satisfying $\slashed{p}_1u(p_1)=mu(p_1)$, $\bar v(p_2)\slashed{p}_2=-m\bar v(p_2)$, and $g_L$ the $U(1)_L$ coupling assumed to be independent of $n$\footnote{The gauge coupling of the $n$-th KK excitation is given by $g_L(n)=g_Le^{-cn^2\frac{M_L^2}{M_s^2}}$, with $c$ a positive numerical constant. When $M_L\ll M_s$, as it is the case in the large extra dimension scenario considered in this paper, the exponential is of order $1$ for all $n\simlt\frac{M_s}{M_L}$, and the gauge coupling can indeed be taken constant. For higher KK modes with $n\gg\frac{M_s}{M_L}$, one has to consider the exponential suppression of $g_L$ and the analysis would be modified. For the values $M_s\sim 10~{\rm TeV}$ and $M_L\sim 10~{\rm GeV}$ that will be considered below, the result presented in this section is thus valid for the first ${\cal O}(10^3)$ KK excitations.}.

The total tree-level amplitude $\mathcal{M}$ for the process $\chi\bar{\chi}\rightarrow l\bar l$ is given by the sum over the KK modes of the vector boson mediator, $\mathcal{M}=\sum_{n=1}^{\infty}\mathcal{M}_n$. Averaging $|\mathcal{M}|^2$ over the incoming helicities and summing over the outgoing helicities of the fermions, one gets
\begin{equation}
\overline{|\mathcal{M}|^2}=2g_L^4\left|\sum_{n=1}^{\infty}\frac{1}{s-n^2M_L^2}\right|^2(t^2+u^2+4s(m_{\chi}^2+m_l^2)-2(m_{\chi}^2+m_l^2)^2),
\end{equation}
where we have defined $\overline{|\mathcal{M}|^2}\equiv\frac{1}{4}\sum_{{\rm Spins}}|\mathcal{M}|^2$, $m_{\chi}$ and $m_l$ are the DM and lepton masses respectively, and $s$, $t$ and $u$ are the Mandelstam variables defined by
\begin{equation}
s=(p_1+p_2)^2,\quad t=(p_1-p_3)^2, \quad u=(p_1-p_4)^2.
\end{equation}
In the center of mass frame, writing the $4$-momentums $p_1=(E,\vec{p_i})$, $p_2=(E,-\vec{p_i})$, $p_3=(E,\vec{p_f})$, $p_4=(E,-\vec{p_f})$ with $E=\sqrt s/2$, $\theta$ the angle between the incoming $\vec{p_i}$ and outgoing $\vec{p_f}$, and neglecting the lepton mass $m_l$, we get: 
\begin{equation}\label{eq:averaged_squared_amplitude}
\overline{|\mathcal{M}|^2}=g_L^4\left|\sum_{n=1}^{\infty}\frac{1}{1-n^2\frac{M_L^2}{s}}\right|^2\left[1+\cos^2\theta+\frac{4m_{\chi}^2}{s}(1-\cos^2\theta)\right].
\end{equation}
We can then compute the differential cross-section of the DM annihilation process, given in the center of mass frame for $2\rightarrow 2$ particle scattering by:
\begin{equation}
\frac{d\sigma}{d\Omega}=\frac{1}{64\pi^2s}\frac{|\vec{p_f}|}{|\vec{p_i}|}\overline{|\mathcal{M}|^2}.
\end{equation}
Explicitly writting the sum in Eq. \eqref{eq:averaged_squared_amplitude} in terms of the cotangent function and using $|\vec{p_i}|=\sqrt{E^2-m_{\chi}^2}$, $|\vec{p_f}|=E$, one gets:
\begin{equation}\label{eq:diff_cross_section}
\frac{d\sigma}{d\Omega}=\frac{g_L^4}{64\pi^2 s}\frac{1}{\sqrt{1-\frac{4m_{\chi}^2}{s}}}\left(-\frac{1}{2}+\frac{\pi\sqrt{s}}{2M_L}{\rm cot}\left(\frac{\pi\sqrt{s}}{M_L}\right)\right)^2\left[1+\cos^2\theta+\frac{4m_{\chi}^2}{s}(1-\cos^2\theta)\right].
\end{equation}
In the non-relativistic limit, $s$ can be expressed in terms of the relative velocity $v_r$ of the annihilating particles as
\begin{equation}
s=4m_{\chi}^2+m_{\chi}^2v_r^2+\mathcal{O}(v_r^4).
\end{equation}
Expanding the expression \eqref{eq:diff_cross_section} in terms of $v_r$ yields:
\begin{equation}
\frac{d\sigma}{d\Omega}=\frac{g_L^4}{256\pi^2 m_{\chi}^2}\left(-1+\frac{2m_{\chi}\pi}{M_L}{\rm cot}\left(\frac{2m_{\chi}\pi}{M_L}\right)\right)^2\frac{1}{v_r}+\mathcal{O}(v_r).
\end{equation}
At lowest order in $v_r$, the differential cross section is thus independent of $\theta$ so that the total annihilation cross section reads:
\begin{equation}\label{eq:annihilation_cross_section}
\sigma v_r=\frac{g_L^4}{64\pi m_{\chi}^2}\left(-1+\frac{2m_{\chi}\pi}{M_L}{\rm cot}\left(\frac{2m_{\chi}\pi}{M_L}\right)\right)^2.	
\end{equation}

\begin{figure}[!h]
        \centering
        \includegraphics[height=70mm]{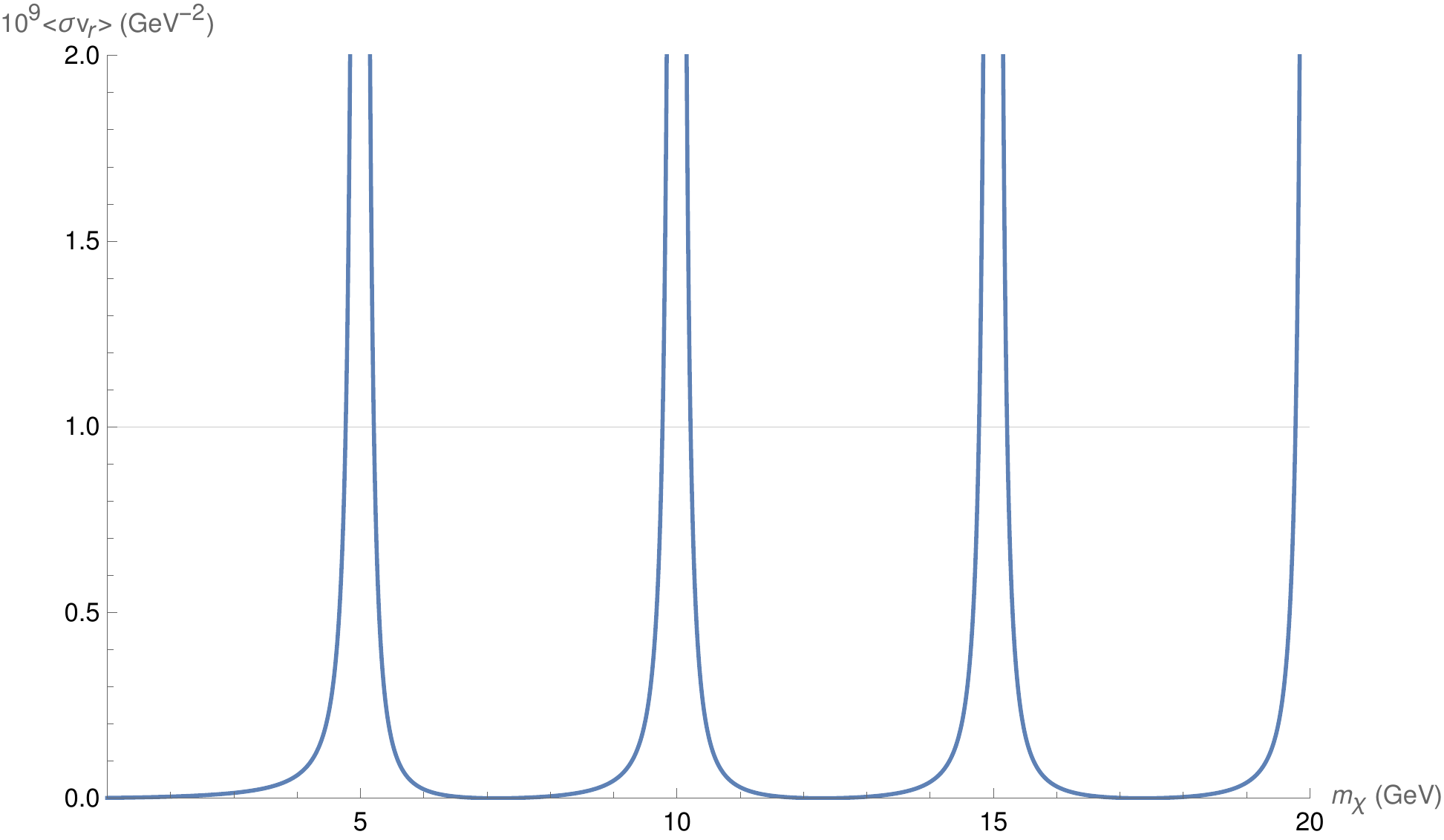}
    \caption{Annihilation cross section $10^9\times\Braket{\sigma v_r}({\rm GeV}^{-2})$ in terms of the DM mass $m_{\chi}({\rm GeV})$, for a compactification scale $M_L\sim 10~{\rm GeV}$ and a coupling $g_L\sim 10^{-2}$.}
    \label{fig:annih_cross_section}
\end{figure}

For a string scale $M_s\sim 10~{\rm TeV}$, $M_L$ and $g_L$ must be of order $M_L\sim 10~{\rm GeV}$ and $g_L\sim 10^{-2}$ in order to accommodate the muon anomalous magnetic moment discrepancy \cite{LA_IA_3}. For such values, the annihilation cross-section \eqref{eq:annihilation_cross_section} is plotted in terms of the DM mass $m_{\chi}$ in Figure \ref{fig:annih_cross_section}. The horizontal grey line indicates the annihilation cross section which yields the observed value for the DM relic density, given by\footnote{The annihilation cross section $\sigma v_r$ given in Eq. \eqref{eq:annihilation_cross_section} being independent of $v_r$, the thermal average $\Braket{\sigma v_r}$ entering in the expression of the DM relic density is trivial in our case : $\Braket{\sigma v_r}=\sigma v_r$.}
\begin{equation}
\Omega h^2\sim \frac{10^{-26}~{\rm cm}^{3}\cdot{\rm s}^{-1}}{\Braket{\sigma v_r}}\sim 0.1\times\frac{10^{-9}~{\rm GeV}^{-2}}{\Braket{\sigma v_r}}.
\end{equation}
The cross-section diverges for all $m_{\chi}=n\frac{M_L}{2}$, and is regulated taking into account the width of the KK resonances, by replacing the vector boson propagator $\frac{1}{s-M_n^2}$ by $\frac{1}{s-M_n^2+i\Gamma_nM_n}$, with $\Gamma_n$ the decay rate of $L_{\mu}^{(n)}$. We thus have to check how this regularisation modifies the function $\Braket{\sigma v_r}$ and in particular if it brings the local maxima of the cross-section below the critical value $10^9\times\Braket{\sigma v_r}=1~{\rm GeV}^{-2}$. For the lowest KK modes $n$, the width of $L_{\mu}^{(n)}$ is dominated by decays into SM leptons, with the corresponding decay rate given by
\begin{subequations}
\begin{align}
\Gamma_n(L_{\mu}^{(n)}\rightarrow l\bar l)&=\frac{g_L^2}{4\pi}M_n\left(1+2\frac{m_l^2}{M_n^2}\right)\sqrt{1-4\frac{m_l^2}{M_n^2}}\Theta(M_n-2m_l),\\
&=\frac{g_L^2}{4\pi}M_n\Theta(M_n-2m_l)+\mathcal{O}(m_l^4).
\end{align}
\end{subequations}
For $g_L\sim 10^{-2}$, we get $\Gamma_n(L_{\mu}^{(n)}\rightarrow l\bar l)\sim 10^{-5}M_n$, so that $\Gamma_n$ can indeed be neglected for the lowest $n$. For higher KK modes, other decay channels contribute to the width, such as the decay of $L_{\mu}^{(n)}$ into lighter KK states, and the analysis would be modified. Focusing on the lightest modes from now on, one thus gets from the plot of Figure \ref{fig:annih_cross_section} that the correct DM relic density is obtained for several values of the $\chi$-fermion mass $m_{\chi}$ centred around integer multiple of $\frac{M_L}{2}$. 
For $M_L\sim 10~{\rm GeV}$, the two first lightest possible DM masses are in tension with the phenomenological constraints coming from dwarf galaxies $\gamma$-ray and CMB observations, which yield a lower bound on the DM mass around $10-15~{\rm GeV}$ \cite{dwarf_galaxies_constraints,CMB_constraints}. In our example, these constraints are thus satisfied for DM masses centred around $m_{\chi}=n\frac{M_L}{2}, n\simgt 3$. Obviously, the constraints are automatically satisfied for $M_L\simgt 30$ GeV.

Since direct couplings between the DM $\chi$ and the SM particles are forbidden, no Yukawa couplings between $\chi$ and the SM leptons are allowed. In model A introduced in Section \ref{sect:models}, this means that among the four possible configurations for $\chi$, the one \allowbreak $\chi(\boldsymbol{1},\boldsymbol{1};0,0,0,1,-1)_{0}$ is not allowed. For the three other quantum number assignments, namely $\chi(\boldsymbol{1},\boldsymbol{1};0,0,0,-1,1)_{0}$, $\chi(\boldsymbol{1},\boldsymbol{1};0,0,0,1,1)_{0}$ and $\chi(\boldsymbol{1},\boldsymbol{1};0,0,0,-1,-1)_{0}$, the Yukawa coupling $L\chi H_u$ is forbidden, and the mass of $\chi$ can arise for instance from brane separation, when the $U(1)_L$ and $U(1)^{'}$ branes are localised at two different points in the extra (large) dimensions of the gravitational bulk.

Obviously, a global (top-down) construction of a fully consistent string model may require the presence of additional branes. Such (hidden) branes or/and corresponding ``messenger'' states may provide alternative DM candidates besides the minimal possibility discussed above. 

\section{Lepton flavour non-universality and the muon $g-2$}
\label{sect:LFNU}
Implicit in the construction carried out above was the assumption of lepton flavour universality, namely that the three families of charged leptons are identical copies of each other (appart from the mass) and that the lepton number gauge boson $L_{\mu}$ couples with the same strength to each of them. Another possibility to address the discrepancy of the anomalous magnetic moment of the muon would be to gauge only the muonic lepton number $U(1)_{L^{(\mu)}}$ replacing the $L$-brane in Figure~\ref{fig:pict_rep} and identifying the leptons of the first and third generation by open strings that do not end on the $L^{(\mu)}$ brane. They could for instance end on a 6-th brane that gauges $L^{(e)}+L^{(\tau)}$, or end on the $U(1)'$ brane in the minimal case, breaking the  total lepton number. As a result, this construction leads to lepton flavour non-universality (LFNU) but its main achievement is to avoid LEP and LHC bounds while still use light KK-excitations of the $U(1)_{L^{(\mu)}}$ gauge boson $L^{(\mu)}_\mu$ in order to account for the $(g-2)_{\mu}$ discrepancy. Their contribution to the muon vertex correction is given by:
\begin{equation}\label{eq:LFNU_contrib}
\Delta a_{\mu}=\sum_n\frac{g_{L^{(\mu)}}^2}{12\pi^2}\frac{m_{\mu}^2}{M_n^2},
\end{equation}
where $m_{\mu}$ is the muon mass, $g_{L^{(\mu)}}$ the gauge coupling of the $U(1)_{L^{(\mu)}}$ and $M_n$ the mass of the $n$th KK excitation of $L_{\mu}^{(\mu)}$. Since $L_{\mu}^{(\mu)}$ does not couple to electrons, its coupling and KK masses evade the LEP bounds and are thus now completely unconstrained.

As mentioned above, a LFNU model can easily be obtained in the framework built in this paper, by replacing the $U(1)_L$ brane by a muonic $U(1)_{L^{(\mu)}}$ associated to a gauge boson $L_{\mu}^{(\mu)}$ with corresponding gauge couplings $g_{L^{(\mu)}}$. Assuming again that $U(1)_{L^{(\mu)}}$ extends along one large extra dimension with a compactification scale $M_{L^{(\mu)}}$, we have $g_{L^{(\mu)}}^2=g_s\frac{M_{L^{(\mu)}}}{M_s}$, with $g_s$ the string coupling, and $M_n=nM_{L^{(\mu)}}$, so that the contribution \eqref{eq:LFNU_contrib} reads:
\begin{equation}
\Delta a_{\mu}=\frac{g_sm_{\mu}^2}{72M_{L^{(\mu)}}M_s}.
\end{equation}
Such contribution can therefore accomodate the discrepancy \eqref{eq:discrepancy} for a compactification scale and a string scale satisfying
\begin{equation}
M_{L^{(\mu)}}M_s\sim g_s\times 5 \times 10^4~{\rm GeV}^2.
\end{equation}
For a string scale $M_s=10~{\rm TeV}$, we thus get a compactification scale $M_{L^{(\mu)}}\sim g_s\times 5~{\rm GeV}$. In the perturbative regime where $\frac{g_s}{4\pi}\simlt 1$, $M_{L^{(\mu)}}$ can therefore vary from the $\mathcal{O}({\rm GeV})$ to $\mathcal{O}(10^2~{\rm GeV})$. Let us note that if $L_{\mu}^{(\mu)}$ also propagates along some extra smaller dimensions with a size larger but near the string length, $g_s$ is suppressed by the volume of these dimensions, further increasing the range of possible values for the compactification scale $M_{L^{(\mu)}}$.

\section{Conclusion}
\label{sect:conclusion}
The Brookhaven National Laboratory experiment E821 together with the recent Muon $g-2$ experiment at Fermilab have pushed the discrepancy between the measured value of the muon anomalous magnetic moment and its Standard Model prediction to $4.7\sigma$, providing a strong hint of new physics beyond the SM. This discrepancy can be explained in the framework of low mass scale strings and large extra dimensions, assuming that the SM lepton number global symmetry (or even the muonic lepton number) is gauged and that the associated gauge boson propagates along (at least) one large extra dimension, so that the main contribution to $(g-2)_{\mu}$ is due to the exchange of its lightest Kaluza-Klein excitations. The work carried out here realised this proposal, by building the minimal embedding of the Standard Model into intersecting D-brane configurations with a gauged lepton number associated to a $U(1)_L$ brane extended along one large extra dimension and which does not participate to the hypercharge combination, as required for phenomenological reasons. Consistency of the models requires the introduction of a fifth brane in a way that all SM mixed anomalies cancel.

The presence of the two extra branes, beyond the SM ones, allows
to introduce in a simple way the right handed neutrino as well as a Dark Matter candidate. For a string scale $M_s\simgt 10~{\rm TeV}$, the bulk of these models exhibits an interesting non-homogeneous structure, with one large extra dimension with a compactification scale  in the range of $\mathcal{O}(10-10^2~{\rm GeV})$ required to explain the $(g-2)_{\mu}$ discrepancy, and several larger extra dimensions with an average compactification scale $\simlt\mathcal{O}(0.1~{\rm GeV})$ in order to lower the string scale in the $\mathcal{O}(10~{\rm TeV})$ region. Within this framework, the anomalous magnetic moment of the muon may provide a hint for the low mass scale strings proposal accessible in future high energy particle colliders.


\acknowledgments
We thank Marco Cirelli and Shaaban Khalil for useful discussions regarding the Dark Matter analysis, as well as Emilian Dudas, Elias Kiritsis and Carlo Angelantonj for enlightening interchanges regarding the D-branes construction.




\end{document}